\documentclass[aps,twocolumn,floats,showpacs]{revtex4}

\usepackage{amsmath,amssymb,graphicx}
\usepackage{epsfig}
\usepackage{epstopdf}
\usepackage{graphicx}
\usepackage{enumerate}

\begin{document}

\title
{Electronic correlation in nanoscale junctions: 
Comparison of the GW approximation to a numerically exact solution
of the single-impurity Anderson model}

\author{X. Wang$^{(1)}$, C. D. Spataru$^{(2)}$, M. S.  Hybertsen$^{(3)}$ and A. J. Millis$^{(1)}$}
\affiliation{$^{(1)}$Department of Physics, Columbia University, New York, NY 10027 \\
$^{(2)} $Center for Electron Transport in Molecular Nanostructures and Center for Integrated Science and Engineering,
Columbia University, New York, NY 10027 \\
$^{(3)}$ Center for Functional Nanomaterials,
Brookhaven National Laboratory, Upton, NY 11973 }

\begin{abstract}
The impact of electronic correlation in nanoscale junctions, e.g. formed
by single molecules, is analyzed using the single-impurity Anderson model.
Numerically exact Quantum Monte Carlo calculations are performed to
map out the orbital filling, linear response conductance and spectral function
as a function of the Coulomb interaction strength and the impurity level position.
These numerical results form a benchmark against which approximate, 
but more broadly applicable,
approaches to include electronic correlation in transport can be compared.
As an example, the self consistent GW approximation has been implemented
for the Anderson model and the results compared to this benchmark.
For weak coupling or for level positions such that the impurity is either
nearly empty or nearly full, the GW approximation is found to be accurate.
However, for intermediate or strong coupling, the GW approximation does not
properly represent the impact of spin or charge fluctuations.
Neither the spectral function nor the linear response conductance
are accurately given
across the Coulomb blockade plateau and well into the mixed valence regimes.
\end{abstract}

\maketitle

\section{Introduction}

Electrical circuits containing nanoscale junctions
are central to nanoscience and condensed matter physics.
The importance comes both from the conceptual issues 
they raise and the possibility of qualitatively smaller electronic devices
with new electrical properties \cite{Nitzan03}. 
Examples include metallic atomic-scale point contacts 
that exhibit quantized conductance \cite{Olesen94} and 
organic molecules linking conducting leads. 
These latter may   form non-resonant tunnel junctions \cite{Salomon03, Tao06} or
single molecule devices whose
conductance is controlled by a Kondo resonance \cite{Park02, Liang02, Yu04, Natelson06}.

The challenge to theory is twofold.
First, the atomic scale specifics of chemical bonding and local structure
can profoundly influence local potentials and energy alignments for the electronic
states that control conduction. Second, even if electron-electron interactions in the leads are well
screened and may be effectively incorporated into the energy bands, 
interactions on the molecule are typically not small, and may 
strongly affect the conductance and the spectrum. 

The important role played by the chemical and structural details 
has led to a strong emphasis in the literature on self consistent theories, 
often utilizing approximate implementations 
of Density Functional Theory (DFT) \cite{Nitzan03}.
In these theories the atomic scale potential and the local bonding structure are
treated in detail while the conductance is calculated via  a Landauer
approach based on the electronic
states derived from the self consistent  Hamiltonian \cite{Datta95, Nitzan01}.
However, these approaches treat excited states in a mean field manner and  there remain 
significant questions concerning the
role of electron correlations and whether the important energy levels are
accurately represented by the mean field theories that are utilized \cite{Neaton06}.

The self consistent DFT approach has proved to be relatively accurate for
metallic point contacts and certain molecular junctions with conductance 
near $G_0 = 2e^2/h$, the quantum of conductance\cite{Nielsen02, Smit02, Thygesen05}.
However, detailed comparison between experiment and self consistent calculations of conductance through
single molecule junctions 
has generally shown a large discrepency, 
up to several orders of magnitude 
in the non-resonant tunneling regime \cite{Nitzan03, Heurich02, Stokbro03}.
Typically the measured conductance is smaller than the calculated conductance.
Unfortunately, these comparisons are complicated by substantial variability
in the measured values for the same molecule
and signficant uncertainty about the atomic scale structure
of the junction near the single molecule link \cite{Salomon03}.
Calculations are typically performed for relatively idealized junction structures
and the conductance can be sensitive to the local geometry for
widely used thiol linkages \cite{Stokbro03, Tomfohr04, Basch05}.

The impact of uncertainties in the junction structure on the comparison
(i.e. of not performing the calculations for the relevant bonding configuration) has recently been clarified,
following the discovery that amine linked junctions produce
single molecule junctions with reproducible conductance measurements, a result
that was understood to derive from a selective bonding motif \cite{Venkataraman06}.
A study of the impact of amine-gold link structures on junction conductance
for benzenediamine gave strong support to the selectivity of the bonding and
showed good agreement between theory and experiment 
for the distribution of conductance \cite{Quek07}. 
However, the magnitude of the calculated conductance exceeded the measured
value by a factor of seven. 
This suggests that even after the junction structure is reliably accounted
for, discrepancies remain and points to the importance of correlation effects beyond the the commonly used  DFT 
based self consistent approach.

These challenges have stimulated theoretical research 
along a number of lines, including implementing self interaction
corrections  \cite{Toher05,Ke07, Toher07} in order to obtain improved
estimates of energy level alignment, correlated basis function
techniques \cite{Delaney04,Muralidharan06} to improve the description of 
the electronic wave functions, fundamental analyses
of the application of DFT to electronic transport to go beyond the Landauer  approach
while remaining within the DFT framework
\cite{Sai05,Koentopp06}, and finally diagrammatic
perturbation analyses of beyond-DFT correlation effects
\cite{Ferretti05a, Ferretti05b, Darancet07, Thygesen07a,Thygesen07b}.

In this paper we focus on the last issue, namely the dynamical consequences
of the on-molecule interactions. 
To discuss the essential physics, we restrict attention to a single resonance
coupled to metallic leads and including the local Coulomb interaction, 
the single-impurity Anderson model \cite{Anderson}. 
We do not consider many body effects
associated with interactions in the leads or between the molecule and the leads. 
The present work examines equilibrium properties
and conductance in the linear response regime only. 

We present a numerically exact quantum Monte Carlo solution to this simple model
as a benchmark against which other approximate approaches can be compared.
As an example of such a comparison, 
we analyse the GW approximation for the electron self energy \cite{Hedin65, Hedin69}.
The GW approach \cite{Hedin65, Hedin69} has been  applied 
in full detail to succesfully predict
the quasiparticle energies for a wide range of solids, surfaces, molecules
and nanosystems \cite{Hybertsen86, Aryasetiawan, Aulbur00, Stan06, Schilfgaarde06}.
Motivated by these successes, 
and noting that it is a conserving approximation \cite{Baym61, Baym62}, 
several authors have begun to apply the 
GW approximation to calculate the electronic properties of
single-molecule conductors\cite{Darancet07,Thygesen07a, Thygesen07b}.
However, it is far from clear under what circumstances the GW approximation
will accurately treat the local correlations and the resulting transport phenomena.

The rest of this paper is organized as follows: 
Section II presents
the model, section III describes the methods (touching on the issue
of the proper definition of the GW approximation for a local orbital), 
section IV presents results for the level filling, conductance, and spectral function, section V discuss the GW approximation
and section VI is a conclusion. An Appendix presents some details of the derivation of
the GW equations we use.

\section{Model}

We study the simplest possible model of a molecular  junction: a
single level which may hold 0, 1 or 2 electrons, has an interaction term
which controls the energy of the two electron state, and is coupled to
an electronic continuum which represents the leads and is taken
to be noninteracting. This is the Anderson impurity model
\cite{Anderson}, represented by the Hamiltonian
\begin{eqnarray}
H&=&\sum_{k\sigma}\varepsilon_k c_{k\sigma}^{\dagger}c_{k\sigma} +
\sum_{\sigma}\varepsilon_d d_{\sigma}^{\dagger}d_\sigma
\\
&&+\sum_{k\sigma} V_k (d^{\dagger}_\sigma c_{k\sigma} + d_\sigma
c_{k\sigma}^{\dagger})+ U n_{d \uparrow} n_{d \downarrow} .
\nonumber \label{AM}
\end{eqnarray}
Here $d^\dagger_\sigma$ creates an electron of spin $\sigma$ on
the localized level  (energy $\varepsilon_d$) and the $U$ term describes
the $d-d$ interaction. $c^{\dagger}_{k\sigma}$ creates an electron of spin $\sigma$
in the lead state with energy $\varepsilon_k$. Because
we will be concerned only with equilibrium
properties, a restriction to a single electronic continuum is 
possible: in a two-lead situation one combination of lead states
decouples from the problem and  the state created by $c^\dagger$ really refers to an electron
in the appropriate ``hybridizing" linear combination.  $V_k$ describes the hybridization
between the level and the lead. 

The crucial quantity that describes the lead electrons is the hybridization function
$\Delta(\varepsilon)=\pi \sum_k |V_k|^2 \delta(\varepsilon-\varepsilon_k)$. 
In our work, we assume a semicircular density of states and a $k$-independent $V$:
\begin{equation}
\Delta(\varepsilon)=V^2\frac{\sqrt{4 t^2-\varepsilon^2}}{2 t^2},\ \ \
|\varepsilon|<2t .
\end{equation}
We choose parameters so that $\Delta(\varepsilon=0)\ll t$
but our conclusions do not depend in any important way of this assumption.

We shall be interested in correlations of the d-electrons, in particular the retarded 
Green's function\cite{Mahan},
\begin{equation}
G_{d\sigma}(\omega)=-i \int_0^\infty\mathrm{d}t \ e^{i(\omega+i0^+)t}\langle[d_\sigma(t), d_\sigma^\dagger(0)]\rangle ,
\end{equation}
from which we obtain the spectral function (index ``d'' is
dropped):
\begin{equation}
A_\sigma(\omega)=-\frac{1}{\pi}\ \mathrm{Im}G_\sigma(\omega) .
\end{equation}
The d-occupancy $\langle n_{\sigma} \rangle$ is given by
\begin{equation}
\langle n_{\sigma} \rangle=\int d\omega
A_\sigma(\omega)f(\omega) .
\label{n}
\end{equation}
Here the Fermi function $f(\omega)=1/(\exp(\beta\omega)+1)$ and 
we have chosen the zero of energy  such that the 
chemical potential $\mu=0$.
The linear response conductance $\sigma$
is given by
\cite{Meir}:
\begin{equation}
\sigma=\frac{e^2}{\hbar}\sum_\sigma\int\mathrm{d}\omega \left [ -\frac{\partial f(\omega)}{\partial\omega} \right ] \frac{\Delta(\omega)}{2}A_\sigma(\omega) .
\label{conductance}
\end{equation}

The non-interacting ($U=0$) model can be solved exactly\cite{Anderson, Mahan}; we obtain
\begin{equation}
G(\omega)|_{U=0}={{\cal
G}_0}(\omega)=\frac{1}{\omega-\varepsilon_d-\Sigma^{V}(\omega)} 
\end{equation}
with the lead self energy
\begin{equation}
\Sigma^{V}(\omega)=\sum_k\frac{|V_k|^2}{\omega-\varepsilon_k+i0^+} .
\end{equation}
For the semicircular density of states the  lead self energy has a simple analytical form (although
attention must be paid to the branch cut structure). We have, on the real and imaginary frequency
axes respectively
\begin{eqnarray}
\Sigma^V(\omega)&=&-i\Delta(\omega)\Theta(2t-|\omega|) 
\nonumber \\
&&+V^2\ \frac{\omega-{\rm sgn} \omega\Theta(|\omega|-2t)\sqrt{\omega^2-4 t^2}}{2t^2} ,\\
\Sigma^V(i\omega_n)&=&V^2\frac{i\omega_n-i{\rm sgn}(\omega_n)\sqrt{(\omega_n)^2+4 t^2}}{2t^2} .
\end{eqnarray}
For $U\neq 0$ the model is no longer analytically solvable. 
The effect of the many-body interaction is expressed mathematically by the 
self energy $\Sigma^U(\omega)$, defined by the relation
\begin{equation}
G(\omega)
=\frac{1}{\omega - \varepsilon_d -\Sigma^V(\omega)-\Sigma^U(\omega)} .
\label{fullG}
\end{equation}

We now qualitatively discuss the behaviour of the model. 
If we assume that the 
the hybridization is weak $(V\ll t)$ 
and the energy range we are considering is well inside the band 
$(2t \gg |\omega|)$, we can take $\Delta(\varepsilon)=\Delta$ $(= V^2/t$ in our case) and 
neglect the real part  of $\Sigma^V$ so that  $\Sigma^V=-i\Delta$. 
The important parameter is $U/\Delta$.
For $U/\Delta\rightarrow 0$ the occupancy varies smoothly with $\varepsilon_d$ 
and the spectral function has a single, approximately
Lorentzian, peak   centered at $\varepsilon_d$ with half-width $\Delta$:
\begin{equation}
A_\sigma(\omega)=\frac{1}{\pi}\frac{\Delta}{(\omega-\varepsilon_d)^2+\Delta^2} .
\end{equation}

For $\Delta\rightarrow 0$ ($U/\Delta\rightarrow\infty$)
 we have an isolated ion decoupled from the leads. There are four states:  the empty state $|0\rangle$ with
energy $E=0$, the  fully occupied state
$|\!\!\uparrow\downarrow\rangle$ with energy $E=2 \varepsilon_d+U$ 
and a magnetic doublet $|\!\!\uparrow\rangle$ or
$|\!\!\downarrow\rangle$ with energy $E=\varepsilon_d$. The $T=0$
spectral function depends on the occupancy: If $\varepsilon_d>0$, the
ground state is $|0\rangle$ and the spectral function consists of an 
addition peak at $\omega=\varepsilon_d$. If $\varepsilon_d<0$ but
$\varepsilon_d+U>0$ the ground state is one of $|\!\!\uparrow\rangle$
or $|\!\!\downarrow\rangle$ and the spectral function has a
removal peak at $\omega=\varepsilon_d<0$ and an addition peak at
$\omega=\varepsilon_d+U>0$. Finally, if $\varepsilon_d+U<0$ the ground
state is $|\!\!\uparrow\downarrow\rangle$ and the spectral
function has only a removal peak, centered at
$\omega=\varepsilon_d+U<0$.

These elementary considerations suggest that  (provided the
d-level occupancy is neither zero nor two) there exists a critical
$U_c$ at which the single-peaked spectral function characteristic
of small $U/\Delta$ changes to the multi-peaked form found in the
large $U$ approximation. A reasonable estimate for the relevant
$U$-scale is provided by the Hartree-Fock (HF)
approximation\cite{Anderson, Coleman} which, for the model used here
yields $U_c/\Delta=\pi$ at occupancy $n=1$.
In the Hartree-Fock approximation the interaction term $U n_{\uparrow}
n_{\downarrow}$ is approximated by $U \langle n_{\uparrow} \rangle
n_{\downarrow} + U n_{\uparrow} \langle n_{\downarrow} \rangle - U
\langle n_{\uparrow} \rangle \langle n_{\downarrow} \rangle$
implying $\Sigma_\sigma^U=U \langle n_{-\sigma}\rangle$ so that
\begin{equation}
G_{\sigma}(\omega)=\frac{1}{\omega-(\varepsilon_d+U\langle
n_{-\sigma} \rangle)-\Sigma^V(\omega)}
\end{equation}
with $\langle n_{-\sigma}\rangle$ fixed from Eq.(\ref{n}).

The Hartree-Fock approximation incorrectly predicts 
that for $U>U_c$ the ground state is spin polarized. 
Corrections to the Hartree-Fock approximation allow the spin to fluctuate, 
leading to the Kondo effect\cite{Kondo, Wilson1975}.
The ground state is non-magnetic, 
characterized by a Kondo energy scale given approximately by\cite{Haldane1978}
\begin{equation}
T_k \approx  0.2\sqrt{2\Delta
U}\exp[\pi\varepsilon_d(\varepsilon_d+U)/(2\Delta U)] \label{Tk} .
\end{equation}
Equation \eqref{Tk} is valid only if $\varepsilon_d(\varepsilon_d+U)<0$. 
Qualitatively, Eq.~\eqref{Tk} shows that the Kondo temperature $T_k$ 
is minimal at the half filling point $\varepsilon_d=-U/2$.

The Kondo ground state is a Fermi liquid, for which the low frequency behavior of the 
many-body self energy is
\begin{equation}
\Sigma^U(\omega)=U\langle n\rangle +\Sigma_0 + (1-Z^{-1})\omega + \Theta(\omega^2, T^2) .
\label{sigmaUFL}
\end{equation}
Here the $U\langle n\rangle$ is the Hartree shift in the d-level energy and $\Sigma_0$ is
any extra chemical potential shift arising from interactions beyond Hartree-Fock.
An important consequence of Eq.~\eqref{sigmaUFL} is that at sufficiently low temperatures
\begin{equation}
A(\omega=0)=\frac{1}{\pi}\frac{\Delta(\omega=0)}{\varepsilon_d^{*2}+\Delta(\omega=0)^2}
\end{equation}
with $\varepsilon_d^*=\varepsilon_d+{\rm Re}\left(\Sigma^V(\omega=0)+\Sigma^U(\omega=0)\right)$
so that at density $n=1$ ($\varepsilon_d^*=0$), $A(\omega=0)=\frac{1}{\pi\Delta}$ and 
from Eq.~\eqref{conductance}, the conductance $\sigma\rightarrow2e^2/h$.

\section{Methods}

In this section we describe both the GW method and the numerically
exact Quantum Monte Carlo (QMC) method to which we compare it.

\subsection{GW}

In the GW approximation \cite{Hedin65, Hedin69} one defines a screened interaction 
$W$ and approximates the electron self-energy as 
\begin{equation}
\Sigma_\sigma^{U, {\rm GW}}(i\omega_{n})=
-T\sum_{i\nu_{m}} G_{\sigma}(i\omega_{n} -i\nu_{m})W_\sigma(i\nu_{m})
\label{SigmaiGW}
\end{equation}
here written as a function of Matsubara frequency \cite{Benedict}, $T$ stands for temperature in unit of energy.
In the extended solid state problem for which the GW approximation 
was originally introduced, 
$W$ is taken to be the screened Coulomb interaction in the charge channel. In 
the impurity model considered here it is essential to retain the spin channel, which 
controls the low energy physics.  
Care must also be taken to respect the Pauli principle: 
the GW approximation corresponds 
to a partial resummation of the infinite set of diagrams which 
define the theory and one must ensure 
that this partial resummation includes all the diagrams necessary 
to respect antisymmetry.

We rewrite the Hubbard interaction $Un_\uparrow n_\downarrow$ as 
a $2\times 2$ matrix in spin space $\hat{V}$ with components
\begin{equation}
V_{\alpha \beta}=U(1-\delta_{\alpha \beta}) .
\label{spindep}
\end{equation}
An alternative definition $V_{\alpha\beta}=U$ is sometimes used in the literature. 
The two definitions are compared in Appendix.

The screened interaction is derived from the irreducible polarizability $P$
through
\begin{equation}
\hat{W}=\left(\hat{I}-\hat{V}\hat{P}\right)^{-1}\hat{V}.
\label{Wdef}
\end{equation}
In the GW approximation, no vertex corrections are included, so the polarizability is just the Random Phase Approximation bubble,
\begin{equation}
P_\alpha(i\omega_{m})= -T\sum_{i\nu_{n}} G_{\alpha}(i\omega_{m} -
i\nu_{n})G_{\alpha}(i\nu_{n}). \label{SCGW1}
\end{equation}
Explicitly, the screened interaction is then
\begin{equation}
W_\sigma(\Omega_n)=\frac{U^2 P_{-\sigma}(\Omega_n)}{1-U^2P_\sigma(\Omega_n)P_{-\sigma}(\Omega_n)} .
\label{SCGW2}
\end{equation}
For later use we note that the quantity $W$ defined in Eq.~\eqref{Wdef}
may be expressed   \cite{Hedin65,Fetter} as a correlation function which
for a paramagnetic ground state on the imaginary time contour is
\begin{equation}
W_{\sigma,\sigma}(\tau)
=\frac{U^2}{4}\left[\langle T \rho(\tau)\rho(0)\rangle + \langle T \sigma(\tau)\sigma(0)\rangle\right]
\label{Wexact}
\end{equation}
with $\rho=n_\uparrow + n_\downarrow-\langle n_\uparrow + n_\downarrow\rangle$ and $\sigma=n_\uparrow - n_\downarrow$.
Eq.~\eqref{Wexact} will be used below in our analysis of the differences between
the GW approximation and the exact results.

Equations \eqref{SigmaiGW}, \eqref{SCGW1}, \eqref{SCGW2} define a self-consistent set of 
equations which are solved numerically by iteration. All quantities are calculated on a real frequency grid, with a frequency range of $\pm 4 t$ and a frequency spacing as small as $t/10^4$ . Where convergence
issues arise (intermediate coupling and the non-magnetic phase) we 
use Pulay mixing\cite{Thygesen07b,Pulay}.

\subsection{QMC}

A numerically exact solution to the Anderson impurity model may be
obtained using Quantum Monte Carlo techniques. For most of the
results obtained here we used the Hirsch-Fye
method\cite{Hirsch1986, Georges1996}; for some of the lowest
temperature data we used the recently developed continuous time
method\cite{Werner2006a, Werner2006b}. 
The QMC calculations were
mostly performed on  a
parallel computer cluster with  20 dual core 2.2 GHz nodes; a typical point requires about 10
hours of computer time per CPU. For Hirsch-Fye either 256 or 512
time slices were used and the lowest accessible temperature was
$T=0.025$. Convergence was verified by comparing two different
time slices or by comparison to the continuous time method. For the
continuous time method the  perturbation orders were typically
$20\sim80$, but at the lowest $T$ orders up to $\sim 140$ were needed.
We typically use $10^4$ time slices for the lowest
accessible temperature $T\approx0.006$.

The central object in the calculation is the
imaginary time Green's function, related to the spectral function
via
\begin{equation}
G(\tau)=\int_{-\infty}^\infty\mathrm{d}\omega\frac{A(\omega) e^{-\tau\omega}}{1+e^{-\beta\omega}}.
\label{GtauA}
\end{equation}
d-electron density and spin correlation functions were also measured.
Inversion of  Eq.\eqref{GtauA} to  obtain $A(\omega)$ from a
computed $G(\tau)$ is a numerically ill-posed problem. We used the
maximum entropy method\cite{Jarrell1996}; while this method
sometimes produces unphysical feature, no difficulties were
encountered in the results described here. Once $A(\omega)$ is
determined, ${\rm Re }G(\omega)$ is obtained from the
Kramers-Kronig relation and then the self-energy from
Eq.~\eqref{fullG}.

It should be noted that the QMC  method is formulated at $T>0$.
The computational expense increases rapidly as $T\rightarrow 0$,
limiting the temperatures which can be reached.

\begin{figure}
    \centering
    \includegraphics[height=\columnwidth, angle=-90]{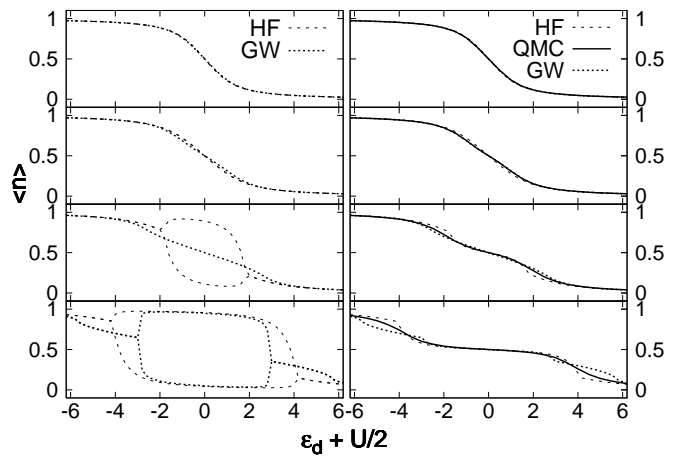}
    \caption{ Comparison of GW, Hartree-Fock and QMC results for d-occupancy. Parameters:
$U=1.05$ (top), 2.1, 4.2, 8.4 (bottom). $V=2.55$, $t=10$,
$\Delta\approx0.65$. Left column: comparison of HF and GW (where
$\langle n_\uparrow \rangle$ and $\langle n_\downarrow \rangle$
differ the two values are shown).
 Right column: comparison of QMC results  to HF and GW results for average density per spin $\langle n_\uparrow+n_\downarrow\rangle/2$.}
    \label{Fig1}
\end{figure}

\section{Results}

\subsection{d-occupancy}
Fig.~\ref{Fig1} shows the d-occupancy as a function of
$\varepsilon_d+U/2$ (the $U/2$ shift puts the particle-hole symmetric
point at zero). The left panel shows the level occupancy obtained from
the HF and GW approximations. The unphysical magnetic solutions
occurring at $U>U_c$ 
are visible as a difference between $\langle n_\uparrow\rangle$
and $\langle n_\downarrow\rangle$. (The QMC calculations yield
$\langle n_\uparrow\rangle = \langle n_\downarrow\rangle$ at all
($U,T$) studied). 
For the parameters studied, $U_c^{\rm HF}\approx 2$.  
The $U_c^{\rm GW}$ depends more strongly 
on temperature than does $U_c^{\rm HF}$ 
making it difficult to determine with precision.
We find $U_c^{\rm GW}\approx 5 $ at $T=0$.  The larger $U_c$ for the GW
approximation arises from the self-consistency (using $G$ rather
than ${\cal G}_0$ to compute $P$);
the renormalization of the Green function suppresses the instability \cite{White92,Suhl67}.

The magnetic phase transition is an artifact of the HF and GW approximation schemes,
but one may expect that the spin-averaged quantities are reasonably reliably represented.
In the right column we plot the
spin averaged $\langle n\rangle$. For strong interaction one can
see clearly the Coulomb blockade plateau. All three methods yield a
roughly correct shape for the occupancy vs level energy curve, in particular giving
approximately correct widths for the Coulomb blockade plateau. At very
weak interaction strength, ($U=1.05$) all three methods agree in detail.
As the interaction is increased, differences appear between the approximate and
exact results. The differences are most pronounced near the edges of the Coulomb
blockade plateau, in the mixed valence regime where charge fluctuations are significant.

\begin{figure}
    \centering
    \includegraphics[height=\columnwidth, angle=-90]{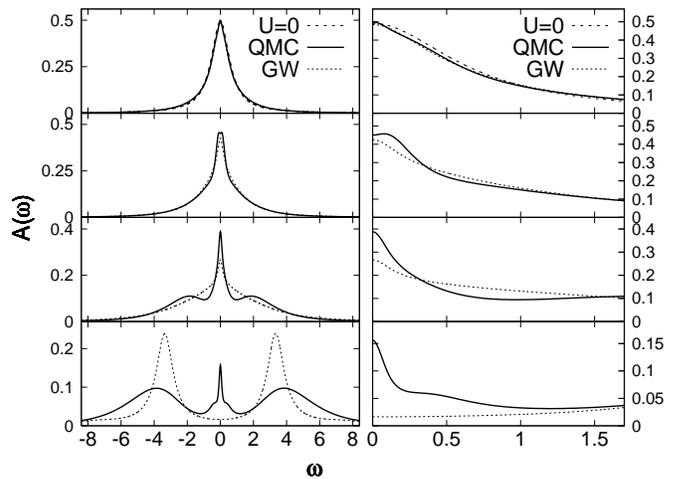}
    \caption{
Electron spectral function calculated by QMC (solid line) 
and GW (dashed line)  at the particle-hole symmetric point
$\varepsilon_d+U/2=0$ for parameters  $T=0.05$, $V=2.55$, $t=10$,
$\Delta\approx0.65$ with $U=1.05$, 2.1, 
4.2, 8.4 (top to bottom panels),  Left panel: wide frequency range;
right panel: expanded view of low frequency range.
For $U=8.4$, GW approximation produces a magnetic solution;
we show the spin-averaged spectral function. 
Due to the fact that $A(\omega)=A(-\omega)$ at half-filling point, only positive frequencies are shown in the right panel.
The non-interacting ($U=0$)  result is also shown as a dashed line
in the top panels.  We believe the very weak dip visible for $\omega$ near $0$ 
in the QMC calculation for $U=2.1$ is an artifact of the analytical continuation procedure.
It is not important for our subsequent discussion.}
    \label{Fig2}
\end{figure}

\subsection{Spectral function}

The computed spectral function $A(\omega)$ is shown in
Fig.~\ref{Fig2}. Focus first on the QMC results over the wide
frequency range (left panels, solid lines). At $U=1.05 \approx U_c^{\rm
HF}/2$ the spectral function is very close to the non-interacting
value. Moving to the middle panel, we see that when $U$ is
increased from $U=2.1 \approx U_c^{\rm HF}$ to $U=4.2$, the Hubbard bands
begin to form, however a central peak remains. At $T=0$ the height of the central peak
should be $1/\pi\Delta\approx0.5$. The reduced height $A(\omega=0)\approx0.4$
is an effect of the non-zero  temperature used in the simulations. For $U=4.2$,
Eq.~\eqref{Tk} implies $T_k\approx0.04$ approximately equal to the
studied $T$. For $U=8.4$ the Hubbard band is well formed, and the
central peak now clearly interpreted as a ``Kondo resonance''
remains. It is interesting that traces of the Kondo resonance are
visible even though the temperature $T=0.05$ studied is much greater than
the Kondo temperature $T_k\approx 0.004$ estimated from
Eq.~\eqref{Tk} as previously noted by Meir {\em et al} \cite{Meir1993}.

The dotted lines in Fig.~\ref{Fig2}  show the results of the GW
approximation. The left panels show that GW agrees reasonably well
with the exact results at $U\le 2.1 \approx U_c^{\rm HF}$, but at
$U=4.2<U_c^{\rm GW}$ the GW does not produce the Hubbard bands and
underestimates the hight of the central peak. At $U=8.4>U_c^{\rm
GW}$ the GW approximation by contrast produces the Hubbard bands
but misses the central peak.
The right column is an expanded view of the central peak. 
For $U=1.05 \approx U_c^{\rm HF}/2$  the two methods agree well with each other,
essentially because the interaction corrections are weak. As the correlations
are increased, differences appear.
We see that even in the $U \approx U_c^{\rm HF}$ regime where the 
GW approximation is reasonably accurate, 
the low frequency lineshape is incorrect, with $A^{\rm GW}$ being 
too low near $\omega=0$ and too high in the wings of the central peak.
The differences become more severe for higher $U$.

\subsection{Conductance}

\begin{figure}
    \centering
    \includegraphics[height=\columnwidth, angle=-90]{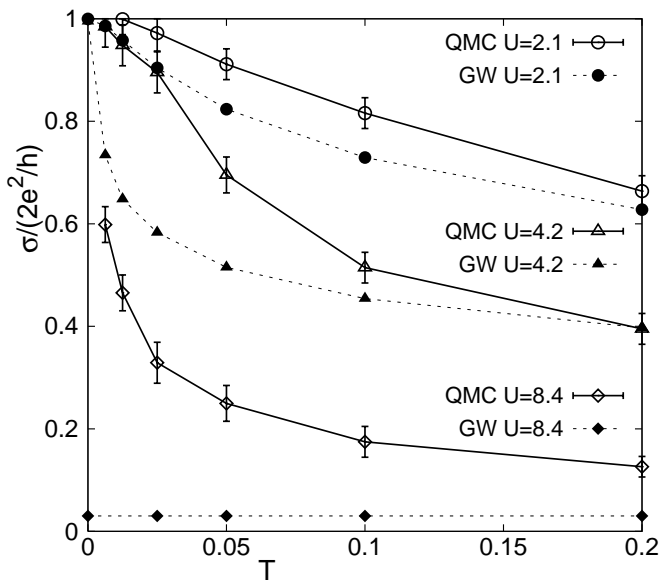}
    \caption{
Linear response conductance as a function of $T$ calculated at half-filling point
$\varepsilon_d+U/2=0$ for different $U$, with 
$V=2.55$, $t=10$, $\Delta\approx0.65$. The Kondo temperatures estimated from 
Eq.~\eqref{Tk} are $T_k\approx 0.09, 0.04, 0.004$ respectively for $U=2.1, 4.2, 8.4$. }
    \label{Fig3T}
\end{figure}

Figure \ref{Fig3T} compares the QMC and GW predictions 
for the linear response conductance 
at  several $U$ values. 
We see that the GW approximation systematically underestimates the conductance, with 
noticeable differences from the QMC values 
even for the smallest $U$-value, $U=2.0 \approx U_c^{\rm HF}$, where the 
GW and QMC spectral functions agree reasonably well. 
We also note that general Fermi-liquid arguments imply 
that as $T\rightarrow 0$, $\sigma=\sigma_{T=0}- T^2/\Theta^2$ with $\Theta$ a temperature scale of order
$T_k$.  However, none of the calculations reveal a clear $T^2$ behavior except for the
QMC calculations at $U=2.1$; we expect that this is because in all of the other cases the Kondo
temperature is close to or below the temperatures studied. 

\begin{figure}
    \centering
    \includegraphics[height=\columnwidth, angle=-90]{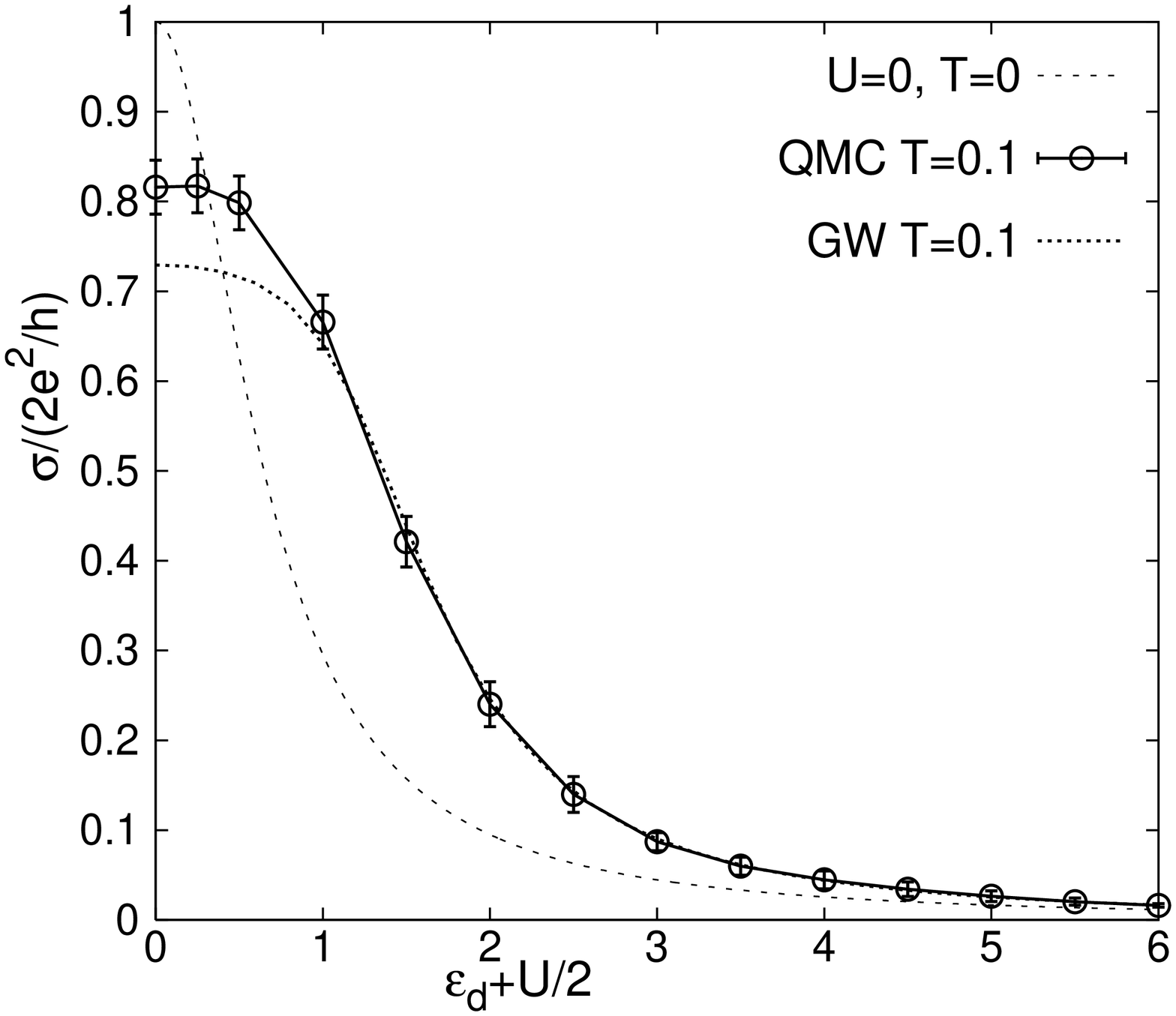}
    \includegraphics[height=\columnwidth, angle=-90]{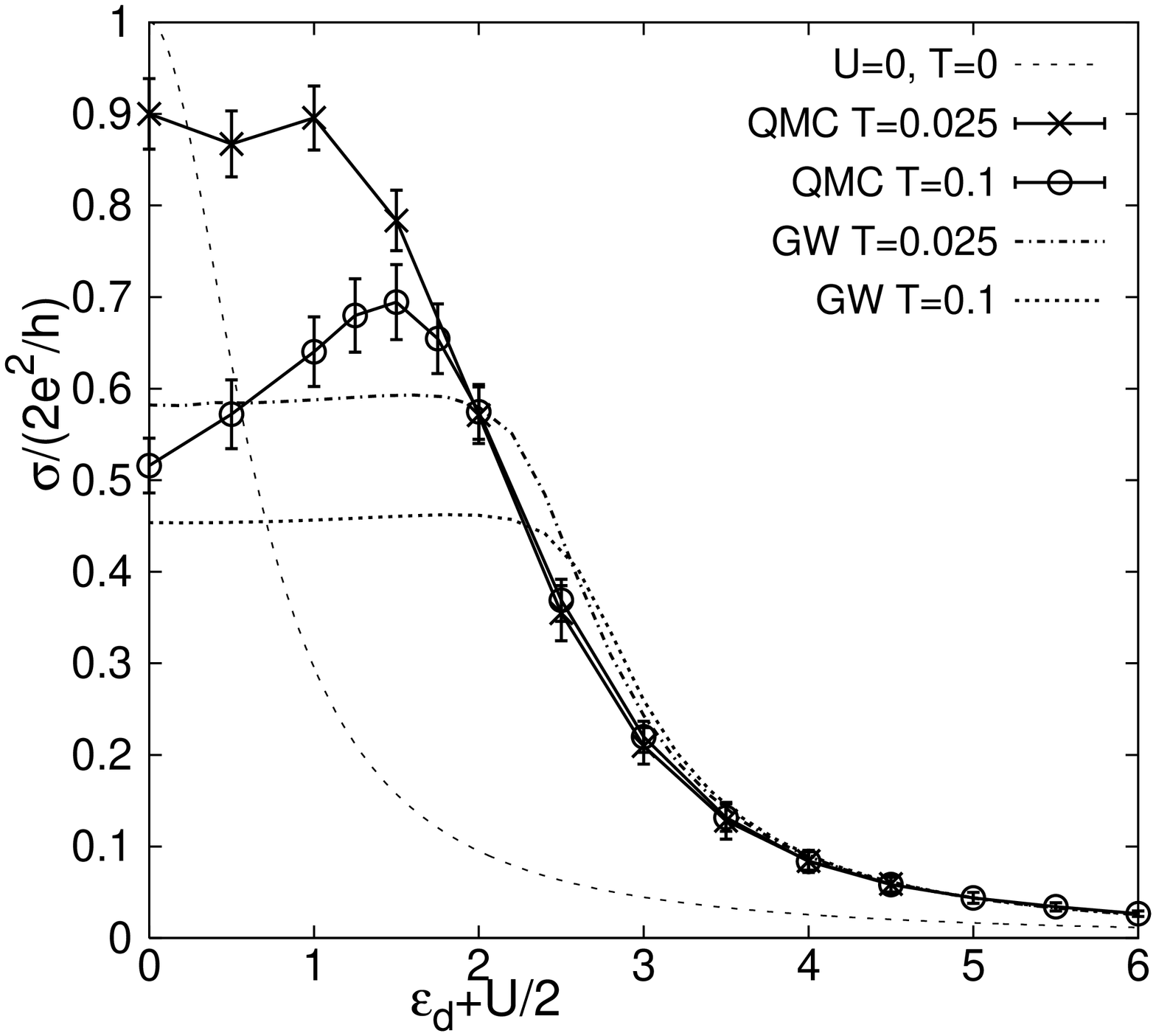}
    \includegraphics[height=\columnwidth, angle=-90]{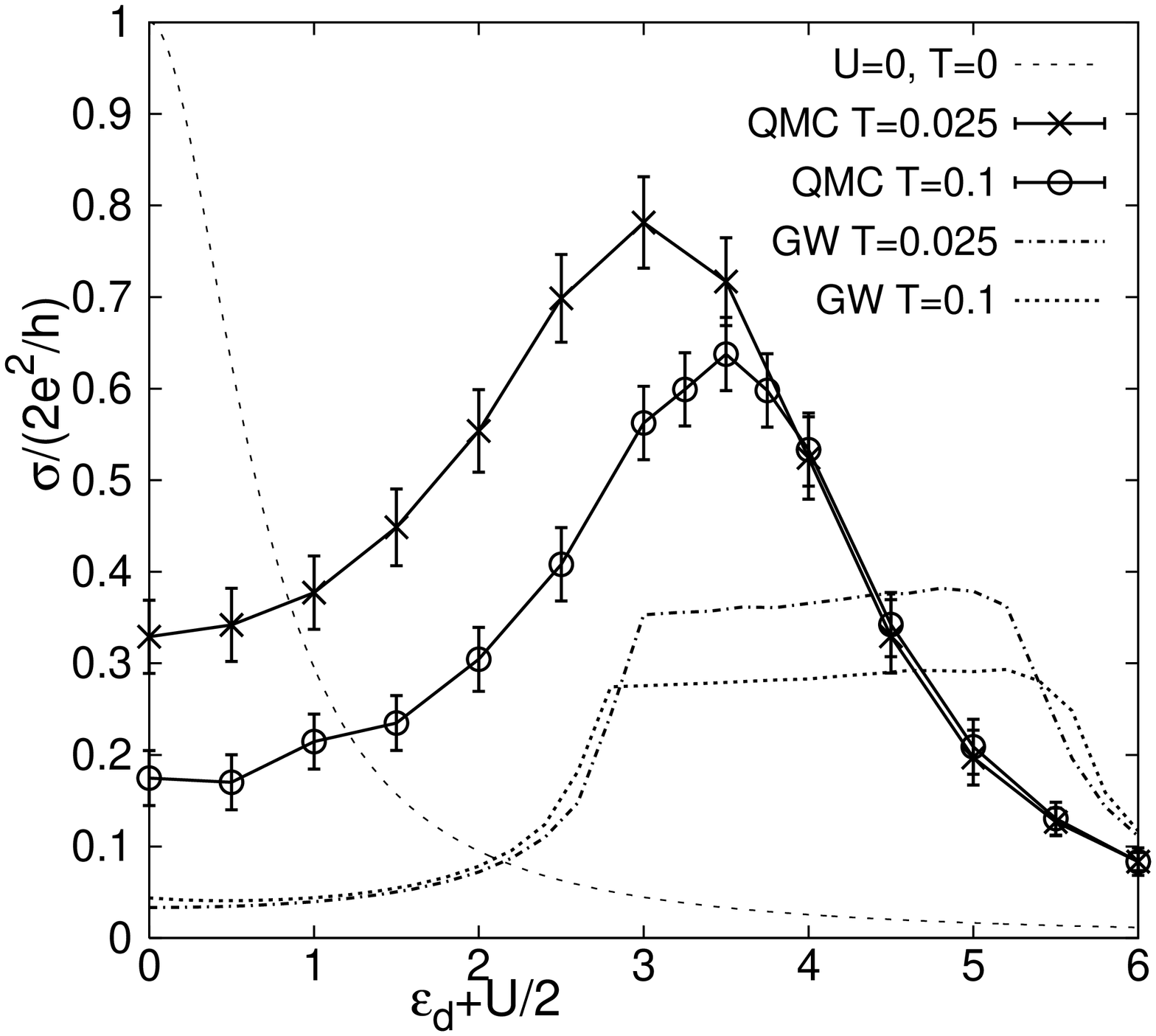}
    \caption{
Linear response conductance as a function of level position $\varepsilon_d+U/2$
calculated for 
$U=2.1$ (upper panel) , $U=4.2$  and $U=8.4$ (bottom panel)
with  $V=2.55$, $t=10$, $\Delta\approx0.65$ at temperatures indicated.
The noninteracting $U=0$, $T=0$
result is also shown for comparison. }
    \label{Fig3a}
\end{figure}

The three panels of Fig.~\ref{Fig3a} show the dependence of
$\sigma$ on level position $\varepsilon_d$ at two different temperatures. At $T=0$
we expect an approximately Lorentzian resonance lineshape,
broadened from the non-interacting value by the density-dependent
level-shift encoded in the real part of the self-energy. As $T$ is
increased the conductance decreases; the decrease from the $T=0$
value is a consequence of many-body scattering;
It is expected to be most pronounced at the particle-hole
symmetric point $\varepsilon_d+U/2=0$. This may be seen
mathematically from Eq.~\eqref{Tk} for the Kondo temperature. In
physical terms, the conductance involves valence fluctuation from the
state $n=1$ to $n=0$ or $n=2$; at the half filled point, these states are most widely separated in
energy, so the fluctuations are most easily disrupted by
temperature.

The top panel shows results for $U=2.1 \approx U_c^{\rm HF}$, along with the
$U=0$ curve for comparison. The  increased width of the interacting curve relative to the noninteracting one is evident
as is the approximately Lorenztian form. The GW and QMC results agree in the
wings of the curve, but disagree in the small $\varepsilon_d+U/2$ regime, with the GW approximation
overestimating the suppression of conductance by thermal fluctuations.

The middle panel shows results for $U=4.2 \approx 2U_c^{\rm HF}$ 
at two different temperatures. 
The QMC curves display the theoretically expected evolution with
temperature  and level position. In the wings of the lineshape (say for $|\varepsilon_d+U/2|>2$)
the curves are temperature independent (for the temperatures studied) and have an approximately
Lorentzian decay. In the central region ($|\varepsilon_d+U/2|<1.5$) the $n(\varepsilon_d)$
curves shown in Fig.~\ref{Fig1} indicate the beginning of a Coulomb blockade plateau and we see correspondingly
a strongly temperature dependent suppression of the conductance.  For these parameters the 
Kondo temperature estimated from Eq.~\eqref{Tk} is $\approx 0.04$ for $\varepsilon_d+U/2=0$; we see that 
for our lower temperature $T=0.025 \approx 0.6T_k$ the conductance approaches the
noninteracting value, as expected.  While the GW approximation shows $T$ dependence,
it is too small for $\varepsilon_d+U/2=0$ while extending strong $T$ dependence
too far with respect to the level position. 
The shape of the $\varepsilon_d$-dependence of the conductance for a given $T$ 
is generally wrong through the Coulomb blockade region.

Finally, the lowest panel shows results for the strongest coupling,
$U=8.4 \approx 4U_c^{\rm HF}$. Reference to
Fig.~\ref{Fig1} shows that for this interaction strength, 
the Coulomb blockade plateau is well formed. 
The theoretically
estimated Kondo temperature at $\varepsilon_d+U/2=0$ is $\approx 0.004$
rather lower than the lowest temperature studied;  correspondingly
the QMC conductance in the Coulomb blockade regime is small and 
strongly temperature dependent. 
In this regime the GW approximation predicts 
a magnetic state with a gap at the Fermi energy and no Kondo resonance, 
so that the conductance at small $\varepsilon_d+U/2$ is qualitatively incorrect.
The GW approximation produces the correct scale 
of $\varepsilon_d+U/2$ at which conductance is restored 
(because it produces a Coulomb blockade plateau of the correct width) 
but gives an incorrect description of the details of the conductance
as a function of level position until beyond the edge of the plateau.

\subsection{Self energy}

\begin{figure}
    \centering
    \includegraphics[height=\columnwidth, angle=-90]{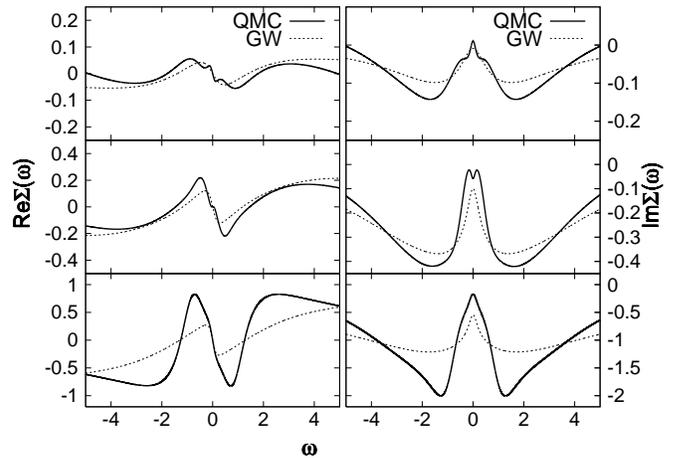}
    \caption{
Real (left column) and imaginary (right column) parts of real axis
self energy calculated using QMC (solid lines) and GW (dashed lines)
at the  particle-hole symmetric point $\varepsilon_d+U/2=0$ for parameters
$V=2.55$, $t=10$, $\Delta\approx0.65$, $T=0.05$  and $U$=1.05 (top),
2.1, 4.2(bottom). The Hartree shift $U\langle n\rangle$ is subtracted from $\mathrm{Re}\Sigma$. Note different vertical axes on left and right panels. }
    \label{Fig4}
\end{figure}

Figure \ref{Fig4} shows the real
frequency behavior of the self energy.  We observe that the
low frequency part of $\mathrm{Re}\Sigma(\omega)$ is linear with a
negative slope, which is consistent with Eq.~(\ref{sigmaUFL}), and
the statement that $Z < 1$. QMC and GW give essentially the same
values of $Z$ for all interaction strengths shown.  
The weak structure visible for the smallest $U$ near
$\omega=0$ is a numerical  artifact of the analytical continuation. 
At higher frequencies
differences between GW and QMC are evident.
In particular, the high frequency tail of the GW curve 
disagrees with the exact analytical result (see Fig.~\ref{Fig6} below).  
Turning now to the imaginary
part of the self energy we first note that the low frequency part of
$\mathrm{Im}\Sigma(\omega)$ is approximately quadratic, which is consistent with
Eq.(\ref{sigmaUFL}). 
Again the weak structures visible very close to $\omega=0$ are
believed to be artifacts of the analytical continuation procedure. 
The non-zero value at $\omega=0$ is a temperature effect. 
The QMC ${\rm Im}\Sigma(\omega)$ is highly peaked; 
however GW fails to produce these peaks.

\section{Analysis of the GW approximation}

\begin{figure}
    \centering
    \includegraphics[height=\columnwidth, angle=-90]{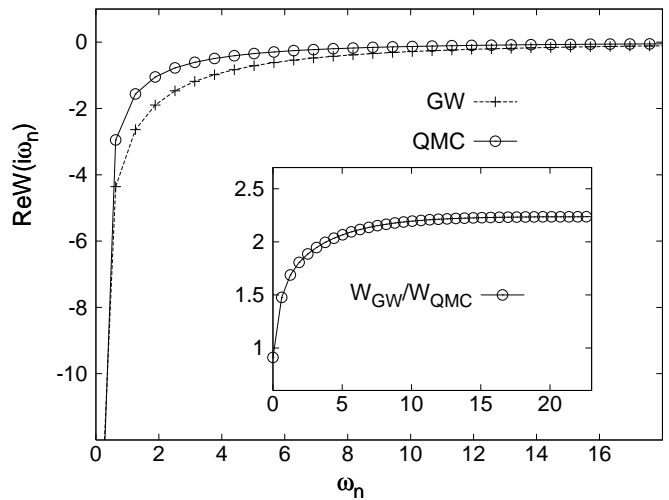}
    \caption{
Comparison of the real part of screened interaction $W(i\omega_n)$ from QMC (Eq. \eqref{Wexact}) and GW 
(Eq. \eqref{SCGW2}) at $\varepsilon_d+U/2=0$. ${\rm Im}W(i\omega_n)=0$. Inset: the ratio $W_{\rm GW}/W_{\rm QMC}$. 
 Parameters: $T=0.1$, $U=4.2$, $\varepsilon_d=-U/2$, $V=2.55$, $t=10$, $\Delta\approx0.65$.}
    \label{Fig7}
\end{figure}

In this section the input to the GW approximation is compared to
the QMC results.  The results provide an explicit measure of the
relative importance of the neglected diagrams for the irreducible
polarizability and the self energy.

Figure \ref{Fig7} compares the screened interaction $W$ obtained from
the GW equation Eq.~\eqref{SCGW2}
to the $W$ obtained from the QMC calculation (Eq.~\eqref{Wexact}) 
for $U=4.2 \approx 2U_c^{\rm HF}$
at the half filling point $\varepsilon_d+U/2=0$.
We see that the interaction strongly increases as $\omega_n\rightarrow 0$;
this is a signature of the slow spin fluctuations which are important to the physics.
The charge fluctuations are suppressed by the Coulomb blockade effect.
Remarkably, the exact and GW results agree very well 
at the lowest frequencies (see inset).
However, the $W$ from the GW approximation is too large 
at intermediate to large frequencies;
screening is incorrectly estimated. 
The ratio $W_{\rm GW}/W_{\rm QMC}$ 
is also shown. The ratio quantifies the effect of vertex corrections, 
which  are neglected in the GW approximation.
The ratio becomes a constant above a certain frequency.

\begin{figure}
    \centering
    \includegraphics[height=\columnwidth, angle=-90]{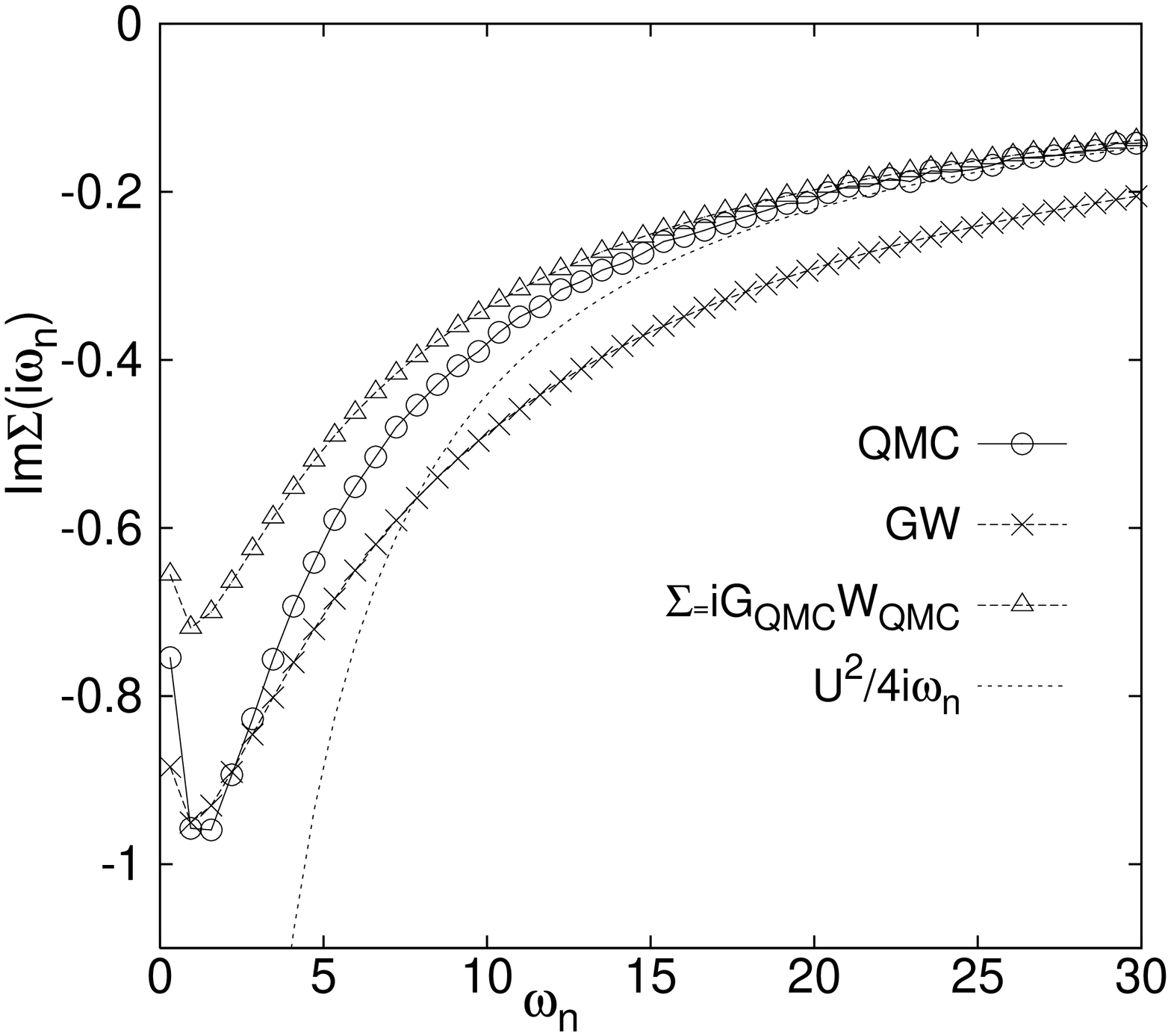}
    \includegraphics[height=\columnwidth, angle=-90]{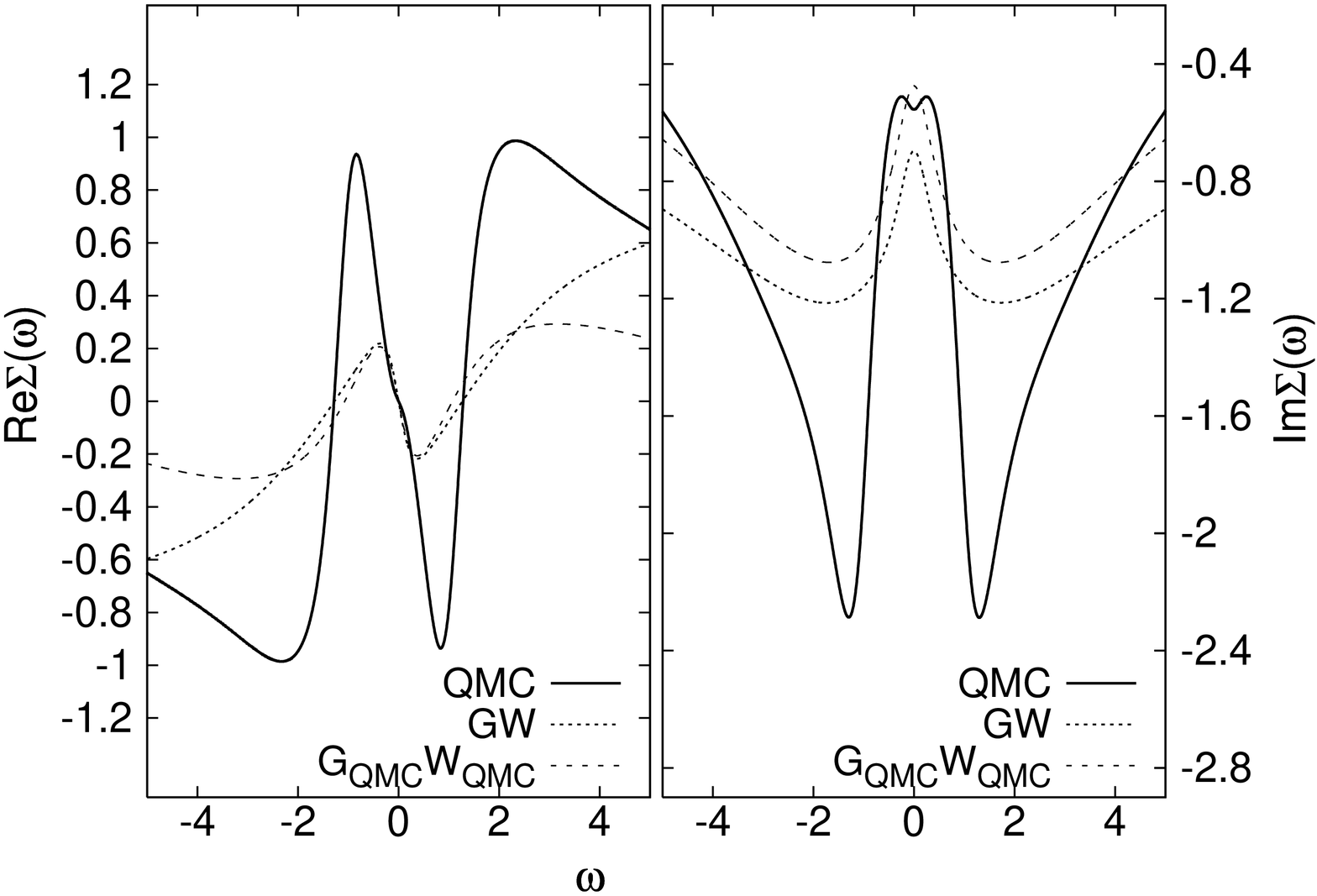}
    \caption{$U$-self-energy on Matsubara axis (upper panel) and 
real axis (lower panels) at half-filling $\varepsilon_d+U/2=0$ computed
using QMC, GW and a hybrid scheme in which the self energy is computed 
according to Eq.~\eqref{SigmaiGW} with the
QMC $G$ and $W$. The analytic asymptotic behavior
$\Sigma(i\omega_n)=U^2/(4i\omega_n)$ is also shown. .
At half filling ${\rm Re}\Sigma(i\omega_n)=U/2$ and $U/2$ is also subtracted from $\mathrm{Re}\Sigma(\omega)$ shown in the lower left panel.  Parameters:
$T=0.1$, $U=4.2$, $V=2.55$, $t=10$, $\Delta\approx0.65$.}
    \label{Fig6}
\end{figure}

\begin{figure}
    \centering
    \includegraphics[height=\columnwidth, angle=-90]{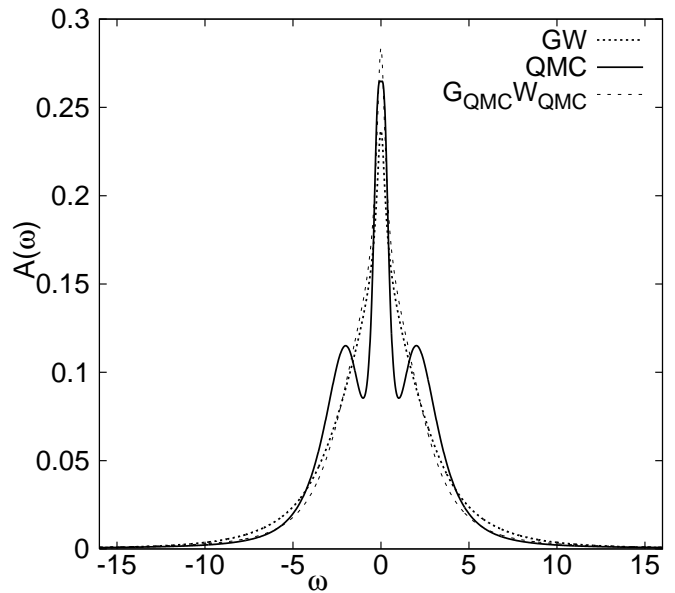}
    \caption{
Comparison of the spectral functions at half-filling point.
$G_{\rm  QMC}W_{\rm QMC}$ curve is the analytical continuation
result from $\Sigma=iG_{\rm  QMC}W_{\rm QMC}$. Parameters: $T=0.1$,
$U=4.2$, $V=2.55$, $t=10$, $\Delta\approx0.65$.}
    \label{Fig6a}
\end{figure}

As shown in the Appendix, the GW approximation gives an incorrect treatment
of the high frequency tail of the self energy, 
implying a violation of the Pauli principle.  
To fix this one may consider replacing the self-consistently determined $G$ and $W$ in
Eq.~\eqref{SigmaiGW} by the exact Green function and polarizibility.
We have used our QMC simulations to measure $G$ and $W$ (from Eq.~\eqref{Wexact}) and used
the results to compute $\Sigma$ from Eq.~\eqref{SigmaiGW}.
Representative  results are shown in
Fig.~\ref{Fig6}. The upper panel shows the imaginary axis self energy.  The GW curve is seen
to have an incorrect asymptotic behavior. Using the QMC $G$ and $W$ produces a curve
with the correct high frequency limit but with an incorrect low frequency behavior. A similar
effect is seen in the lower panels, which display the analytically continued curves.
Both GW and ``$G_\mathrm{QMC}W_\mathrm{QMC}$'' curves substantially underestimate
the frequency dependence of $\Sigma$. Additional insight comes from
the spectral functions shown in Fig \ref{Fig6a}. 
We see that  the ``$G_\mathrm{QMC}W_\mathrm{QMC}$''
curve fails to reproduce the Hubbard band structure, 
and gives an incorrect magnitude
at low frequency.

This comparison explicitly shows that the diagrams neglected in
the GW approximation for the self energy are significant in this regime.
The use of the $W_\mathrm{QMC}$ automatically includes the vertex
corrections for the screened interaction.
Approximations beyond GW also include vertex corrections explicitly in
the expression for the self energy \cite{Hedin65, Hedin69}.
Figures \ref{Fig7}-\ref{Fig6a} highlight the role they play for $U \gtrsim U_c^{\rm HF}$.

\section{Conclusion}

In this paper we used a numerically exact Quantum Monte Carlo method 
to obtain  results for the density, conductance and spectral function 
of the single-impurity Anderson model, a simple theoretical paradigm problem for 
molecular conductors.  
The Anderson  model is characterized by two dimensionless combinations
of  three parameters: 
a level position $\varepsilon_d$, a level width $\Delta$ and an interaction $U$.
In the conventions adopted in this paper, 
the impurity level is half filled when $\varepsilon_d+U/2=0$.
A relevant measure of the interaction strength 
is provided by the Hartree-Fock approximation, which predicts
a magnetic state for interaction strengths greater than a critical value. 
We define $U_c^{\rm HF}$ to be the critical value for
the half filled level.  
For $U>U_c^{\rm HF}$ the impurity spectral function is characterized by
a three-peak structure with upper and lower Hubbard bands at $\omega\approx \pm U/2$ 
and a central ``Kondo" peak which controls the linear response conductance.

Our results are intended as benchmarks against which other, more approximate
but more widely applicable methods may be compared. We compared our results to those obtained from the GW approximation,
a self-consistent partial resummation of diagrammatic perturbation theory.
The GW method is attractive because it can be combined with band theory 
to yield material-specific results, but its efficacy at treating strong correlations is unclear.
Recent literature has argued that the GW method provides
a reasonable description of the physics of the low $T$ ($T=0$ limit) transport properties of molecular conductors for a range of intermediate 
$U>U_c^{\rm HF}$ \cite{Thygesen07a,Thygesen07b}.

We showed that for weak to 
moderate coupling regime ($0 < U < U_c^{\rm HF}$) 
or for a nearly full or nearly empty $d$-level, 
the GW approximation provides a reasonable description.
As the interaction $U$ approaches $U_c^{\rm HF}$, some systematic deviations
are observed in the shape of the spectral function near $\omega = 0$
and the dependence of the conductance on the level position
near $\varepsilon_d+U/2=0$.
For the  intermediate coupling regime ($U_c^{\rm HF}< U < U_c^{\rm GW}$) 
and the  strong coupling regime ($U > U_c^{\rm GW}$), 
the GW appproximation gives 
an inaccurate representation of the spectrum and the linear response conductance
across the Coulomb blockade plateau. 
The accuracy is also limited for the mixed valence regions at the boundaries.

\begin{figure}
    \centering
    \includegraphics[height=\columnwidth]{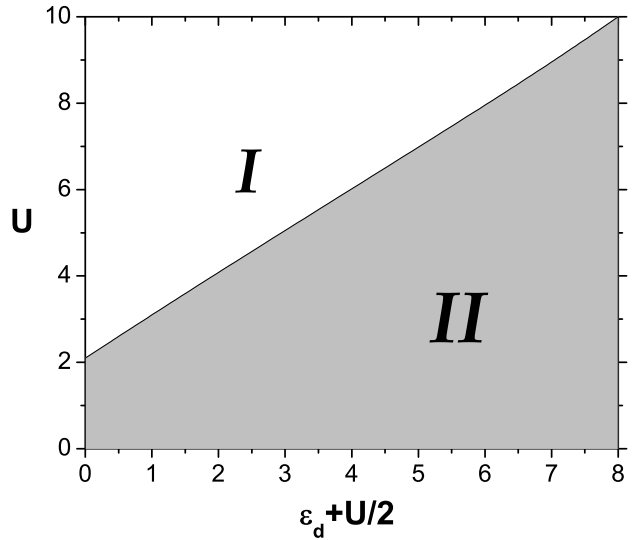}
    \caption{
Phase plane of interaction strength $U$ and bare level energy $\varepsilon_d+U/2$ shows regions where GW approximation works poorly (white, region I) and well (grey, region II). Our criterion for the latter is $|\sigma^{\rm GW}-\sigma^{\rm QMC}|/\sigma^{\rm QMC}<15\%$.
 Parameters: $T=0.1$,
$V=2.55$, $t=10$, $\Delta\approx0.65$.}
    \label{GWvalidity}
\end{figure}

These findings are summarized in Fig.~\ref{GWvalidity}, 
which shows the regions where GW does 
and does not work well in the plane of interaction strength and level position. 
``Works well" is of course an imprecise definition; 
in constructing Fig.~\ref{GWvalidity} we defined 
``works well" as ``GW conductance within 15\% 
of QMC conductance at $T=0.1$". 
The criterion is temperature dependent 
as can be seen from Fig.~\ref{Fig3T} and interaction dependent, as can be seen from 
slope of the boundary  line, which is less than $1/2$. 
We find that the GW approximation is reliable when the level is tuned
so that the density is far enough outside the Coulomb blockade region.  
For $U=4.2$ the GW approximation becomes reasonable for densities
at the edge of the Coulomb blockade plateau, but for  
$U=8.4$ the density must be tuned well away from the plateau before GW becomes accurate.

Qualitatively, in the parameter regime in which GW 
produces a central peak in the spectral function,
it does not produce the Hubbard sidebands, 
while the Hubbard sidebands are produced only as a consequence
of an unphysical magnetic ordering instability.  
Papers in the literature interpret the central peak 
found in the non-magnetic GW calculation
as a Kondo resonance. 
We believe this interpretation is not correct. 
It is interesting to note, however, that some aspects
of the many body physics 
(for example the low frequency ``mass renormalization"  $\partial \Sigma/\partial \omega$
or the spin correlation function) are correctly given by GW. 
This has been seen for the self consistent 
second order self energy as well \cite{White92}.

The self consistent GW approximation has the virtue of being
a conserving approximation \cite{Baym61, Baym62}.
In physical systems where the local molecular levels remain nearly
filled or nearly empty, or where the hybridization is large,
our results show that the GW approximation will be reasonably accurate.
This suggests that an approach based on the GW approximation may
be very useful for molecular conductors in the non-resonant tunneling
regime where large discrepencies exist between theory and experiment.
However, when the local Coulomb interactions on the molecule
are strong, the GW approximation does not accurately represent the
impact of local spin and charge fluctuations.
Neither the spectral distribution nor the linear response conductance are given
properly.
Application of the GW approximation to nanoscale junctions
in the Kondo regime is not well justified.
The analysis of the screened Coulomb interaction $W$ 
and the evaluation of the GW approximation 
for the self energy operator with the the exact (QMC) $G$ and $W$
showed that vertex corrections are quite significant in these cases.
Unfortunately, while there are systematic guidelines for including
vertex corrections properly so as to maintain a conserving approximation,
the resulting theory is substantially more complex \cite{Bickers89}.

Our conclusions are based on linear response. Other situations, in particular the out-of-equilibrium Coulomb Blockade regime remain to be studied. 

{\it Acknowledgements}  We thank
D. Reichman for helpful conversations. XW thanks A. Comanac and
P. Werner for discussions and computer help. CDS thanks K. S. Thygesen
for sharing details about the Pulay mixing implementation.
This work was primarily supported by the 
Nanoscale Science and Engineering
Initiative of the National Science Foundation under
NSF Award Number CHE-0641523 and by the New York
State Office of Science, Technology and Academic Research (NYSTAR). 
This work was partially supported by the 
US Department of Energy, Office of Basic Energy Sciences,
under contract number DE-AC02-98CH10886
and by the National Science Foundation under grant number DMR-0705847. 

\appendix
\renewcommand{\theequation}{A-\arabic{equation}}
\setcounter{equation}{0}
\section*{Appendix}

\begin{figure}
    \centering
    \includegraphics[height=\columnwidth, angle=-90]{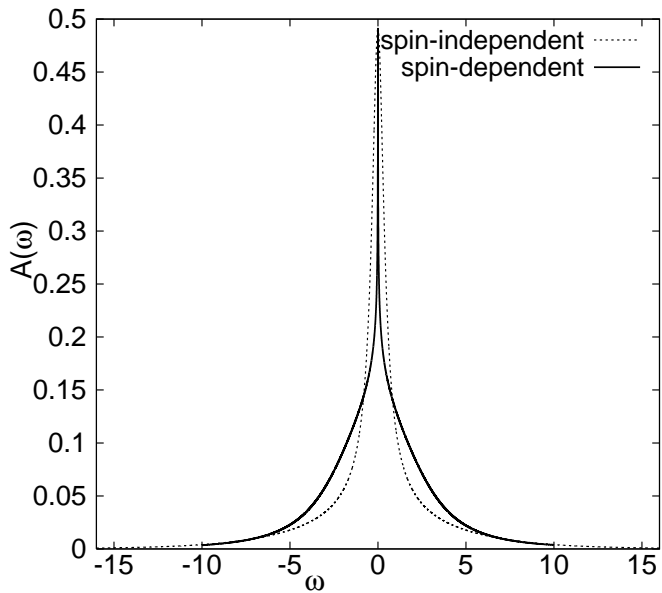}
    \caption{
Comparison of the spin-dependent and spin-independent
approaches to the GW approximation. Parameters used: $T=0$, $U=4.2$,
$V=2.55$, $t=10$, $\Delta\approx0.65$.}
    \label{Fig5}
\end{figure}

This Appendix discusses technical details of the GW calculations.
The local interaction depends on electron spin. One may consider two forms:
\begin{eqnarray}
V_{\alpha\beta}&=&U \hspace{0.65in} \mathrm{``spin-independent"} ,
\label{spin-indep} \\
V_{\alpha \beta}&=&U(1-\delta_{\alpha \beta})\hspace{0.15in} \mathrm{``spin-dependent"} .
\label{spin-dep}
\end{eqnarray}
We now show that the asymptotic high-frequency behavior of the GW self energy implies that
the spin-dependent interaction Eq.~\eqref{spin-dep} is more appropriate than
the spin-independent interaction Eq.~\eqref{spin-indep}.  

It is convenient to separate out the Hartree and Fock terms, writing
$\Sigma=\Sigma^{\rm HF}_\sigma+{\tilde \Sigma}_\sigma^{\rm GW}$.  For the spin-independent interaction the Hartree term for spin $\sigma$
is $U\langle n_\uparrow +n_\downarrow\rangle$ while the Fock term is $- U\langle n_{ \sigma}\rangle$; for the spin dependent
interaction the Fock term vanishes and the Hartree term is $U\langle n_{\bar \sigma}\rangle$. In either case  we have
\begin{eqnarray}
{\tilde \Sigma }_\sigma^{\rm GW}(\omega\rightarrow\infty)
&\approx&
-\frac{1}{\pi\omega}
\int_{0}^{\infty}{\rm d}\varepsilon~{\rm Im}W_\sigma(\varepsilon)
\nonumber \\
&=&-\frac{ {\rm Im}W^{\rm TO}_\sigma(t=0)}{\omega}
\end{eqnarray}
where TO stands for time-ordered. In the exact perturbation theory
analysis, the screened interaction $W$ is related to the spin-spin
correlation function through the polarizability $\Pi$
\cite{Fetter}:
\begin{equation}
\hat{W}^{\rm TO}(t)=\hat{V}\delta(t)+\hat{V}\hat{\Pi}^{\rm TO}(t)\hat{V}
\end{equation}
where:
\begin{equation}
\Pi^{\rm TO}_{\sigma\sigma'}(t)=-i\langle
T\left\{[n_{\sigma}(t)-\langle n_{\sigma}(t)\rangle
][n_{\sigma'}(0)-\langle n_{\sigma'}(0)\rangle ]\right\}\rangle
\end{equation}
In the spin-dependent case, the $(1-\delta_{\alpha,\beta})$ term in the interaction
implies that 
$W_{\sigma,\sigma}$ involves
only the correlator for the opposite spin, so  one finds the
following asymptotic behavior:
\begin{equation}
{\tilde \Sigma}_\sigma^{\rm GW}(\omega\rightarrow\infty)\rightarrow
\frac{U^2}{\omega}\left( \langle n^2_{\bar \sigma}\rangle -\langle n_{\bar \sigma}\rangle^2 \right) .
\label{GWapprox}\end{equation}

On the other hand, for the spin-independent interaction, all spin indices are involved and one obtains
\begin{eqnarray}
{\tilde \Sigma}_\sigma^{\rm GW}(\omega\rightarrow\infty)&\rightarrow&\frac{U^2}{\omega} \sum_{\sigma,\sigma'}
\left(\langle n_{\sigma}n_{\sigma'}\rangle -\langle n_{ \sigma}\rangle \langle n_{ \sigma'}\rangle\right) .
\nonumber
\end{eqnarray}
Thus we see that the spin-dependent interaction reproduces approximately the analytically known asymptotic behavior of the 
self-energy 
${\tilde \Sigma}_\sigma^U(i\omega_n)= 
\frac{U^2}{i\omega_n}[\langle n_{\bar \sigma}\rangle 
( 1- \langle n_{\bar \sigma}\rangle )]$ 
whereas the spin-independent interaction does not.
The asymptotic behavior is only approximately reproduced 
because GW cannot account correctly for 
$\langle n_{\bar\sigma}^2\rangle=\langle n_{\bar\sigma}\rangle$. 
We show in the main text that using the exact $W$ 
yields the correct asymptotic behavior of $\Sigma$
but still does not produce an accurate approximation at general $\omega$.
The spin-independent case provides a much worse approximation,
which would be wrong even if the exact correlations functions were used. 
This is an indication
that the spin-dependent 2-particle interaction is to be prefered
over the spin-independent one in the context of the GW
approximation applied to the Anderson model.

Fig.~\ref{Fig5} shows a comparison of spin-dependent and spin-independent GW at zero temperature. 
Comparison to the lowest temperature QMC result shows that the lineshape calculated from the spin-dependent
GW approximation is  closer to the QMC lineshape than is the result of the spin-independent calculation. 
This is because the spin-dependent approach is free of self-interaction effects and it accounts for some of 
spin-spin quantum fluctuations, whereas the spin-independent approach accounts only for 
density-density quantum fluctuations.

\end{document}